# Modeling Financial Time Series using LSTM with Trainable Initial Hidden States

Jungsik Hwang
*Samsung Research*
*Samsung Electronics Co., Ltd.*
Seoul, South Korea
jungsik.hwang@gmail.com

*Abstract*—Extracting previously unknown patterns and information in time series is central to many real-world applications. In this study, we introduce a novel approach to modeling financial time series using a deep learning model. We use a Long Short-Term Memory (LSTM) network equipped with the trainable initial hidden states. By learning to reconstruct time series, the proposed model can represent high-dimensional time series data with its parameters. An experiment with the Korean stock market data showed that the model was able to capture the relative similarity between a large number of stock prices in its latent space. Besides, the model was also able to predict the future stock trends from the latent space. The proposed method can help to identify relationships among many time series, and it could be applied to financial applications, such as optimizing the investment portfolios.

## I. INTRODUCTION

Time series can be found in many fields, such as finance, economics, engineering, science, etc. Extracting previously unknown patterns and information in time series is central to many real-world applications [1, 2]. Among them, mining financial time series has been extensively studied for many years. Modeling financial time series, especially stock market data, poses several challenges. For instance, time series are usually high dimensional so measuring similarity is challenging [2-4]. Moreover, the dynamics of the stock market is a complex phenomenon and financial time series are very noisy [5, 6]. As a result, predicting stock prices has been known to be notoriously difficult [6, 7].

With the recent advancement of deep learning, many deep learning techniques were introduced for analyzing financial time series [8, 9]. Often, analyzing stock data was preceded by extracting latent features from the financial time series. Convolutional Neural Network (CNN) and Autoencoder (AE) were widely used to extract features from the data [6, 7, 10]. Then, analyzing stock data, such as predicting stock price or trend was performed on the extracted features. Long Short-Term Memory (LSTM) has been a popular choice for modeling financial time series, especially for predicting stock price [11-15]. Other types of deep learning models, such as Generative Adversarial Network (GAN) have been also employed in modeling financial time series [5].

In this study, we introduce a novel approach to model financial time series with a deep neural network model. We propose an LSTM with trainable initial hidden states. Here, trainable initial hidden states indicate that the model optimizes the initial value of the hidden state of LSTM for each sequence in the dataset. In our approach, the model learns to reconstruct stock price changes from the initial states in an unsupervised manner. Reconstructing time series from the Recurrent Neural Network (RNN) equipped with the trainable initial states has been used in previous studies [16-18]. In this paper, we extend the previous works by employing LSTM which has shown superior performance in modeling time series [19]. The proposed model can reconstruct high-dimensional time series data with its parameters. Especially, the trainable initial hidden states play an important role in capturing the similarity/dissimilarity between the time series. Although similarity for time series can be defined in various ways, the proposed model can illustrate the relative similarity between a large number of stock data by learning to reconstruct them from the initial states. In the experiment with KOSPI (Korea Composite Stock Price Index) dataset, it was shown that the proposed model was able to capture temporal similarities among a large number of stock price data in the initial states. In addition, the model was also able to forecast future stock trends from the initial states.

## II. RELATED WORKS

Deep learning has shown remarkable performances in many fields including finance and economics. See [8, 9] for the recent review on deep learning for financial time series. One of the popular applications of deep learning in this field is predicting stock price and trend. Among the deep learning models, LSTM and its variants have been widely favored due to its superior capability of processing time series [11-14]. Often, LSTM was combined with other deep learning models. For example, in [14], the authors proposed a combination of CNN and bi-directional LSTM to forecast stock price for the next seven days. Specifically, CNN was used for feature extraction and bi-directional LSTM was used to analyze temporal data. Similarly, in [15], the authors introduced a combination of wavelet transforms (WT), stacked autoencoders (SAEs), and LSTM for predicting stock price. Especially, the SAEs were used to extract deep features for stock price forecasting. Then, those features were used by LSTM to predict the next day's closing price. Also, in [7], Convolutional Autoencoder (CAE) was used to extract image features from the index price daily data that were represented graphically. Then, LSTM was used to predict the future trend of the index.

In our work, a single deep learning model can learn an abstract representation of time series with its parameters and the model can make predictions from the representation. Unlike previous studies that converted time series into graphical representations [6, 7, 10], the proposed model can directly learn from the raw time series. Therefore, the architecture of the proposed model is simpler than the ones consisting of different deep learning models. In addition, representations that emerged in the initial states of the proposed model can illustrate the relationship among many time series.

## III. LSTM WITH TRAINABLE INITIAL HIDDEN STATES

### A. Architecture

We propose LSTM [20] equipped with trainable initial hidden states (Fig. 1). Here, the "trainable" initial hidden states indicate that the initial values of the hidden states of LSTM ($h_{t=0}$) are optimized for each sequence $k$ during training. This is the key difference between the proposed approach and the conventional approaches. The initial hidden and cell states ($h_{t=0}$, $c_{t=0}$) are required to compute the gates and states in the LSTM layer at the first time step ($t=1$). The common practice is to randomize both initial hidden and cell states. Instead of randomizing the initial states, we optimize the initial hidden states for each sequence $k$ and we set the initial cell state as zero. Previous studies [16-18] have shown that RNN with the trainable initial states was able to embed high-dimensional time series (robot's action) into its parameters. In this study, we employed LSTM since it has shown superior performance in modeling time series with long term dependencies [19, 20].

The proposed model consists of the initial hidden states, the LSTM layer, and the fully connected layer (Fig. 1). The initial hidden states ($h_{t=0}$) are the collection of the initial hidden states for each sequence in the dataset at $t=0$ (1). $N$ is the number of sequences in the dataset.

$$h_{t=0} = \{h_0^1, h_0^2, \ldots, h_0^k, \ldots h_0^N\} \quad (1)$$

**Forward Pass.** When reconstructing the sequence $k$, the model's input at the first step and the initial cell states are set to zero ($x_1^k = c_0^k = 0$). Then, the initial hidden states of the sequence $k$ ($h_0^k$) are given to the LSTM layer. At each time step $t$ (where $t > 0$), the hidden state ($h_t^k$) and the cell state ($c_t^k$) of the LSTM layer are computed as follows.

$$h_t^k = o_t^k \odot tanh(c_t^k) \quad (2)$$

$$c_t^k = f_t^k \odot c_{t-1}^k + i_t^k \odot g_t^k \quad (3)$$

Where $o_t^k, f_t^k, i_t^k, g_t^k$ are the output, forget, input and cell gates respectively. $\odot$ is the Hadamard product. See [21] for the computation of the gates of LSTM. Then, the fully connected layer generates the model's output ($y_t^k$) from the hidden state of the LSTM layer (4). Note that the output is fed back to the input at the next time step (5). This feedback loop is often referred to as the closed-loop generation [16]. In this setting, the model does not require external input, but it can generate sequences solely from the initial states and its own output.

$$y_t^k = Linear(h_t^k) \quad (4)$$

$$x_t^k = \begin{cases} 0 & , when\ t = 1 \\ y_{t-1}^k & , otherwise \end{cases} \quad (5)$$

**Training.** The proposed model is trained in an unsupervised manner where the model is trained to reconstruct time series in the dataset. The reconstruction error for $k^{th}$ sequence ($E^k$) is the difference between the model's output ($y^k$) and the target ($\hat{y}^k$) from the dataset (6). $D$ is the Euclidean Distance function. During training, the reconstruction error is minimized by updating the model's parameters, such as weights, biases, and the initial hidden states. Note that the initial hidden states are trained for each sequence $k$ ($\partial E^k / \partial h_0^k$) in the dataset.

$$E^k = \sum_t D(y_t^k, \hat{y}_t^k) \quad (6)$$

### B. Key Features

The proposed model learns to represent financial time series using its parameters. Especially, the initial hidden states are trained separately for each sequence. The model can reconstruct different stock prices from the different initial states. Consequently, the initial hidden states play a key role in capturing differences between the stock price trends. In other words, the initial hidden states alone are enough to illustrate differences in key properties of the data, without other model parameters. In this sense, the model's initial hidden states work similarly to dimensionality reduction which is important when processing high-dimensional time series [2].

An ability to represent time series in a reduced space (initial hidden states) leads to several benefits. First, the model can illustrate the similarity between a large number of stock price changes. Understanding time series similarity is key to many applications [1-4]. Although it is not straightforward to define a similarity function for time series [3, 4], the relative similarities among stock data can be illustrated from the initial states. More specifically, it is assumed that similar stock price trends would learn similar values of the initial hidden states. For example, the previous study [22] showed that the representations located nearby in the latent space encoded the similar robot behaviors. Second, feature-based algorithms, such as clustering and classification can be performed on the values of the initial states. Previous studies often used different deep learning models for the feature extraction phase and the prediction phase [7, 14, 15]. In our approach, a single model can learn time series representation in the initial states, and also it can make predictions from the initial states. That is, feature extraction and prediction phases are simultaneously performed during training in our work.

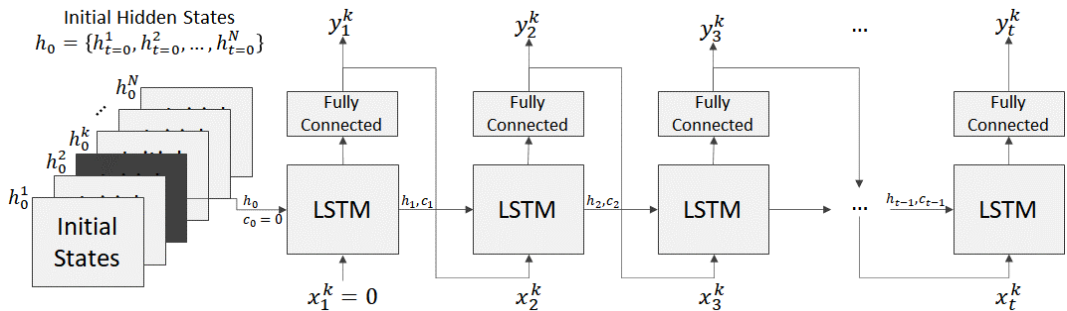

Fig. 1. The model used in this study consists of the initial hidden states, LSTM, and the fully connected layers. The model can generate sequences from the initial states. To generate the $k^{th}$ sequence, the initial hidden states of the $k^{th}$ sequence ($h_{t=0}^k$) is fed to the LSTM layer at the beginning. The initial cell state ($c_{t=0}^k$) and the input ($x_{t=0}^k$) are set to zero. Then, the model's output ($y_{t=0}^k$) is computed from the fully connected layer. At each time step, the model's output is fed back to the input at the next time step. During training, the model's learnable parameters, including weights, biases and the initial hidden states are optimized to minimize the reconstruction error in an unsupervised manner.

## IV. RESULTS

### A. Experiment Settings

**Dataset.** The stock prices of the companies listed in Korea Composite Stock Price Index (KOSPI) from Jan 1st to Dec 31st, 2019 were used. The moving average filter with the window length of five days was applied to the adjusted closing price and some stocks with missing values were excluded. As a result, a total number of 735 stocks were used in our dataset. During training, the stock price change was used as a target ($\hat{y}$). For each sequence $k$, we computed the price change relative to the price of the first day (7). $price_1^k$ is a stock price of a $k^{th}$ sequence at the first day – Jan 1st, 2019.

$$\hat{y}_t^k = \frac{price_t^k - price_1^k}{price_1^k} \qquad (7)$$

**Model Setting.** The model consisted of 128 LSTM cells and 735 initial hidden states. The ADAM optimizer [23] was used and the initial learning rate was set to 0.001. The model was trained for 100,000 epochs. The model was implemented using PyTorch [24] and the code is available online (https://github.com/mulkkyul/lstm-initStates).

### B. Representations in the Initial Hidden States

We first examined the representations of stocks learned in the initial hidden states. We conducted Principal Component Analysis (PCA) on the initial hidden states to visualize them in a 2-dimensional space. Fig. 2 illustrates the initial hidden states ($h_0$) after the training. The colors denote the industrial sector of the company. For clarity, only the companies listed in KOSPI50 are displayed with their names. Although some companies in the same industrial sector appeared nearby (e.g., NCsoft (036570.KS), Naver (035420.KS), and Kakao (035720.KS) on the top center), the representations that emerged in the initial states seemed not relevant to the industrial sector.

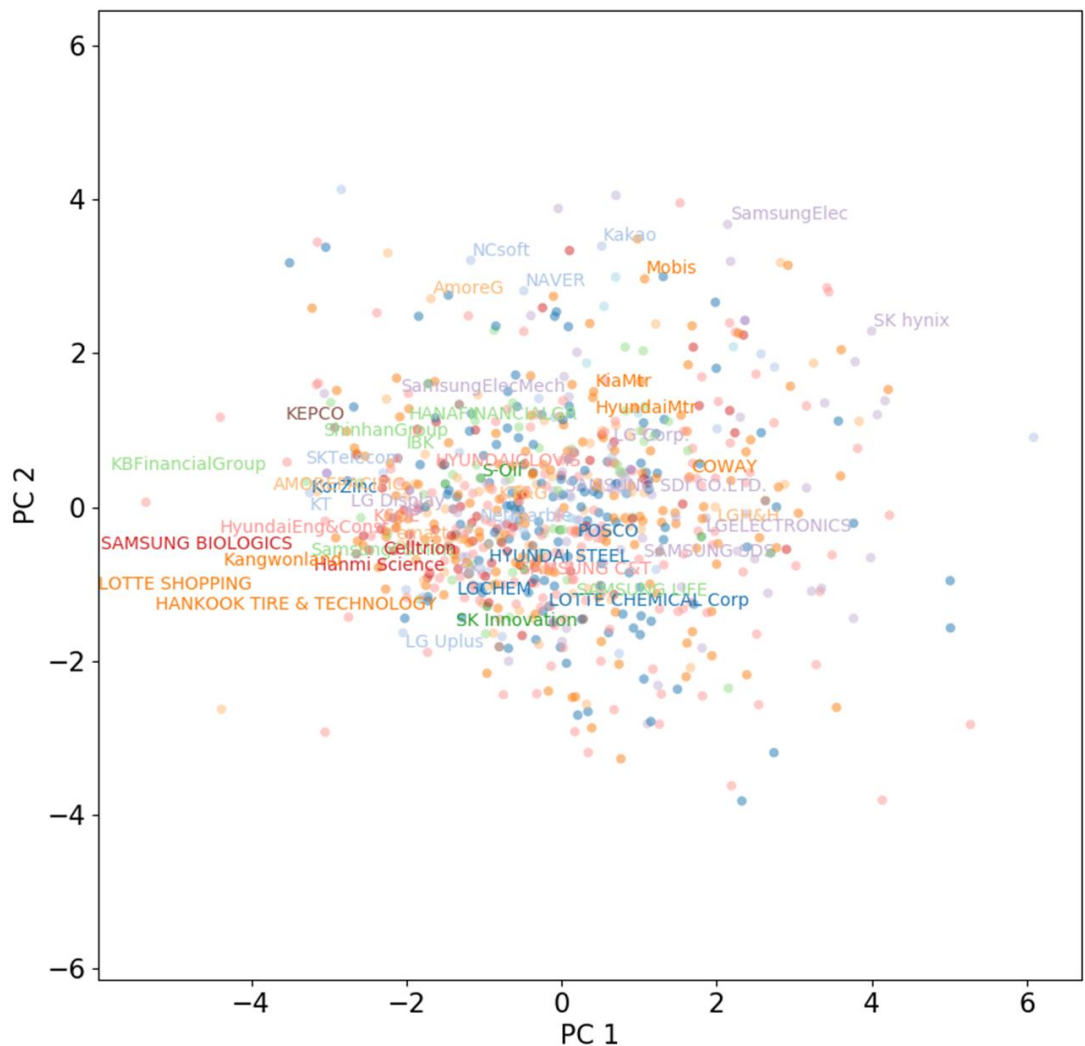

Fig. 2. The initial hidden states ($h_0$) of the model after training. The initial states are visualized with PCA (Principal Component Analysis). The X and Y axis indicate the first and the second principal component respectively. Each dot is a representation of a stock in the dataset (*N = 735*) and the color indicates the company's industrial sector. For clarity, only 50 companies listed in KOSPI50 are displayed with its name.

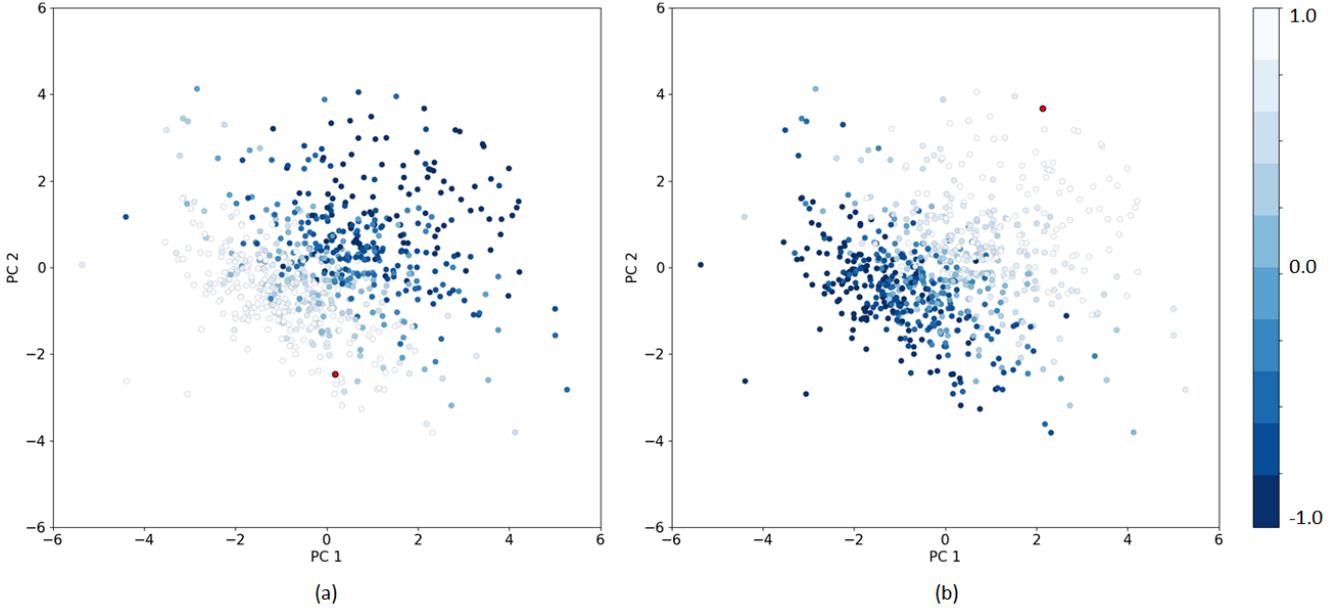

Fig. 3. The initial hidden states colored with respect to the correlation to the target stock. For each plot, a target stock is marked as a red circle and the correlation between the target and the other stocks was used to color the initial states. The color of each dot represents the Pearson's correlation coefficient ranging from -1.0 (dark blue) to 1.0 (light blue). The target company in each subplot is (a) Daegu Department Store (006370.KS) and (b) Samsung Electronics Co., Ltd. (005930.KS).

To further investigate the stock's representations, we colored the initial states with respect to correlation to a target stock (Fig. 3). A correlation has been widely used to measure the similarity between time series [15]. The target stock is marked as a red circle and the Pearson's correlation coefficient to the target stock was used to color the initial states. Both Fig. 3 (a) and (b) show that the representations located near the target tend to be more correlated. As the distance to the target increases in the initial states, correlation to the target decreased. It implies that the relationship among stocks, such as correlation, in the original dimension remains quite similar in the reduced dimension (i.e. initial states).

This highlights the key benefits of the proposed model. Measuring time series similarity has been known to be challenging [3, 4]. In addition to the difficulty of defining the similarity measure, measuring the relative similarity of a large number of time series requires huge computation. Previous studies [6, 7, 10] used another deep learning models, such as CAE and CNN, to learn features that are used for measuring stock similarity. Our finding suggests that the single model with the initial hidden states can be used to measure time series similarity (stock price trend). In addition, the relative similarity among many time series could be easily illustrated by visualizing the initial hidden states.

*C. Predicting Stock Price*

We also investigated the prediction capability of the proposed model. We compared the prediction accuracy of the two cases. In the "short" case, the model learned three months of stock price data (1st Jan to 31st Mar 2019). In the "long" case, the model learned six months of stock price data (1st Jan to 30th Jun 2019). We measured the prediction accuracy of the six subsequent months for each case (1st Apr to 30th Sep for the "short" case, and 1st Jul to 31st Dec for the "long" case). The prediction accuracy was computed in terms of the mean absolute error. For each case, the model predicted the stock prices of the entire companies in the dataset ($N=735$).

Fig. 4 (a) illustrates the prediction accuracy of the two cases (red for the "short" case and green for the "long" case). The horizontal axis indicates time and the vertical axis indicates the mean absolute error. The result showed that the size of the dataset did not have a huge impact on prediction accuracy. Although the short case showed a slightly smaller error than the long case, both cases showed similar error ranges. Both cases showed the low error at the beginning of prediction. The error tends to increase with time, indicating that predicting a stock price in the distant future is harder than the one in the near future. Fig. 4 (b) shows an example of the stock price change of a company best predicted in the "long" case. The figure shows the actual stock price (black dashed line) and the model's output (cyan solid line). The model successfully reconstructed the first six months of stock price (1st Jan to 30th Jun 2019). Also, the model's prediction for the following six months of the stock price was similar to the actual price.

Predicting stock price has been known to be very challenging due to many factors [6]. Although the prediction error increased with time, the result showed that the proposed model can forecast future stock prices from the initial states that encoded the relationship between many numbers of stocks. However, at the same time, it should be noted that the prediction capability of the proposed model is limited since it predicts the stock trends without any external input. As introduced in several previous works [8, 9], utilizing other information than stock prices might help to enhance the model's prediction performance.

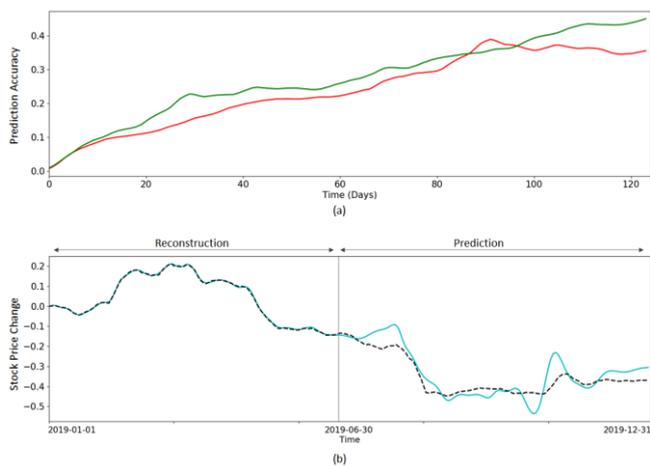

Fig. 4. (a) The accuracy of stock price prediction of two cases: (red) The model trained with a "short" dataset (stock prices from 1st Jan to 31st Mar 2019), and (green) the model trained with a "long" dataset (stock prices from 1st Jan to 30th Jun 2019). The prediction accuracy was measured in terms of mean absolute error. The horizontal axis indicates the six subsequent months for each case (1st Apr to 30th Sep for the "short" case, and 1st Jul to 31st Dec for the "long" case). (b) The stock price change of a company (Cosmax Inc.; 192820.KS) which was best predicted by the model with a "long" dataset. The stock price change is relative to the price of the first day (7). The actual price is in the black dashed line and the predicted one is in cyan solid line.

## V. Conclusion

In this study, we introduced a novel approach to modeling financial time series using a deep learning model. We used the LSTM network with the trainable initial hidden states. By learning to reconstruct stock price change, the model was able to represent those data with its parameters. Especially, the initial hidden states played an essential role in capturing differences among the data. The experiment with KOSPI stock data showed that the model was able to illustrate the similarity between the financial time series by learning the different values in the initial hidden states. By visualizing the initial states, the relative similarity between a large number of stock data could be identified. Therefore, the proposed method can help to identify previously unknown relationships among a large number of data. We also examined the prediction accuracy of the model. Although the prediction accuracy decreased with time, it was shown that the model was able to forecast stock prices from the initial states that encoded the relationship among stocks. The proposed approach could be applied to optimize the investment portfolio consisting of stocks with similar/dissimilar price trends. Also, it can be extended to model high-dimensional time series in other fields since the proposed method can work with sequences with arbitrary lengths and dimensions. In future work, it would be interesting to examine how to regularize the initial states depending on the task.